\documentclass[a4paper,conference]{IEEEtran}

\usepackage{cite}  
\usepackage{graphicx}  
\usepackage{subfigure}
\usepackage{verbatim}
\usepackage{aas_macros}
\usepackage{amsmath}
\usepackage{amssymb}
\usepackage{bm}
\usepackage{Wurster_macros}

\usepackage{fancyheadings}
\newcommand\iBzn[1]{\emph{iB}$_{\text{-}z}^{#1}$}      
\newcommand\nBzn[1]{\emph{nB}$^{#1}_{\text{-}z}$}    
\newcommand\nBzp[1]{\emph{nB}$_{\text{+}z}^{#1}$}   
\newcommand\nBzpn[1]{\emph{nB}$_{\pm z}^{#1}$}     
\newcommand{\tf}{$3\times10^5$}            
\newcommand{\os}{$10^6$}                     
\newcommand{\ts}{$3\times10^6$}           
\newcommand{\ntf}{$N=3\times10^5$}     
\newcommand{\nts}{$N=3\times10^6$}    
\newcommand\citeparen[1]{(\!\!~\cite{#1})}  
\newcommand\Bmax{$B_\text{max}$}     
\newcommand\Bcen{$B_\text{cen}$}      

\begin{document}

\title{Resolving numerical star formation: \\ A cautionary tale }

\author{\IEEEauthorblockN{James Wurster \& Matthew R. Bate}
\IEEEauthorblockA{School of Physics and Astronomy,\\
University of Exeter\\
Exeter, EX4 4QL, United Kingdom\\
j.wurster@exeter.ac.uk}}


\maketitle

\begin{abstract}
Resolution studies of test problems set baselines and help define minimum resolution requirements, however, resolution studies must also be performed on scientific simulations to determine the effect of resolution on the specific scientific results.  We perform a resolution study on the formation of a protostar by modelling the collapse of gas through 14 orders of magnitude in density.   This is done using compressible radiative non-ideal magnetohydrodynamics.  Our suite consists of an ideal magnetohydrodynamics (MHD) model and two non-ideal MHD models, and we test three resolutions for each model.  The resulting structure of the ideal MHD model is approximately independent of resolution, although higher magnetic field strengths are realised in higher resolution models.  The non-ideal MHD models are more dependent on resolution, specifically the magnetic field strength and structure.  Stronger magnetic fields are realised in higher resolution models, and the evolution of detailed structures such as magnetic walls are only resolved in our highest resolution simulation.  In several of the non-ideal MHD models, there is an off-set between the location of the maximum magnetic field strength and the maximum density, which is often obscured or lost at lower resolutions.  Thus, understanding the effects of resolution on numerical star formation is imperative for understanding the formation of a star.
\end{abstract}

\section{Introduction}
When modelling a phenomenon, such as the formation of a protostar, there are many aspects that must be considered.  First, one must consider the relevant and important physical processes that actually describe the phenomenon; these are often selected based upon observational and empirical evidence.  Next, one must chose the numerical algorithms and the resolution; the numerical algorithms include how to model both the physical processes required for realism and the artificial algorithms required for numerical stability.  Even within the smoothed particle hydrodynamics (SPH) framework, there are multiple numerical formalisms for each algorithm, and the final results will differ based upon this choice.  For example, we previously showed the effect that different artificial resistivity algorithms had on the formation of a protostellar disc \cite{Wurster+2017}.  Thus, the choice of numerical algorithms is equally as important and the choice of which physical processes to include.

Choosing a resolution requires a trade-off between accuracy and speed.  There are always minimum criteria that must be resolved, such as the Jeans mass \cite{BateBurkert1997} for molecular clouds, or  the Toomre-mass and scale-height \cite{Nelson2006} for protostellar discs.  However, even if these minimum criteria are initially met, information may still be lost in simulations where certain regions require even higher minimum resolution criteria, or in regions where these criteria become violated due to the adaptive nature of compressible SPH.  The preferable resolution is the convergence limit -- i.e. the minimum resolution simulation to which all higher resolution simulations yield the same results.  However, this convergence limit can be difficult or impossible to determine due to the computational cost of running numerous simulations at increasing resolution.  To date, there are many processes in astrophysics where numerical convergence has not been reached due to computational limitations, including fragmentation of protostellar discs \citeparen{MeruBate2011criteria,MeruBate2011converge,MeruBate2012,Meyer+2018}, turbulence \citeparen{Price2012turb,TriccoPriceFederrath2016,BoothClarke2019}, and also star formation \citeparen{BateTriccoPrice2014,WursterPriceBate2016,WursterBatePrice2018ff}.

Investigating resolution in simplified  tests, such as MHD shock tubes \citeparen{BrioWu1988,RyuJones1995} or the Orszag-Tang vortex \cite{OrszagTang1979} yield useful information and baselines, but it is not until resolution is investigated in practical models that its effect is fully understood.  This is due to the complexity of the practical models where many different physical processes operate simultaneously, unlike in the simplified tests.  Moreover, the resolution effects, and hence the convergence limit, will likely be different for different phenomena.

As a practical investigation of the effects of resolution, we will investigate the formation of a protostar.  To self-consistently model this, we will model the gravitational collapse of gas through 14 orders of magnitude in density; this will be done in the presence of weakly ionised gas \cite{MestelSpitzer1956} and strong magnetic fields \cite{HeilesCrutcher2005}, which are observed in star forming regions.  We include radiation hydrodynamics \cite{BateKeto2015}, magnetohydrodynamics (MHD) \cite{Price2012}, and the non-ideal MHD processes \citeparen{WardleNg1999,BraidingWardle2012accretion,BraidingWardle2012sf}, specifically Ohmic resistivity, ambipolar diffusion and the Hall effect.  Ohmic resistivity and ambipolar diffusion are dissipative terms that weaken the magnetic field, and the Hall effect is a dispersive term that modifies the geometry of the magnetic field.  In non-ideal MHD, the gas is composed of charged gas that is coupled to the magnetic field and neutral gas that is independent of the magnetic field; the neutral gas can slip through the magnetic field, but is influenced by it via collisions with the charged gas.  Non-ideal MHD represents physical resistivity, but we also include artificial resistivity \cite{Phantom2018} in our models for numerical stability.  

We will begin in \secref{sec:num} by describing our methods, followed by our initial conditions in \secref{sec:ics}.  We will investigate the effect of resolution in \secref{sec:results}, and we will briefly discuss and summarise our results in \secref{sec:conc}.

\section{Numerical Methods}
\label{sec:num}
\subsection{Radiation non-ideal magnetohydrodynamics}
The equations of self-gravitating, compressible, radiation non-ideal magnetohydrodynamics are
\begin{eqnarray}
\frac{{\rm d}\rho}{{\rm d}t} & = & -\rho \nabla\cdot \bm{v}, \label{eq:cty} \\
\frac{{\rm d} \bm{v}}{\rm{d} t} & = & -\frac{1}{\rho}\nabla \cdot \left[\left(p+\frac{B^2}{2}\right)\mathbb{I} - \bm{B}\bm{B}\right] \notag \\
 &-& \nabla\Phi + \frac{\kappa \mbox{\boldmath$F$}}{c} + \left.\frac{{\rm d} \bm{v}}{\rm{d} t} \right|_\text{art}, \label{eq:mom} \\
\frac{\rm d}{{\rm d}t} \left(\frac{\bm{B} }{\rho} \right) & = & \left( \frac{\bm{B}}{\rho} \cdot \nabla \right) \bm{v}  \notag \\
   &+& \left.\frac{\rm d}{{\rm d}t} \left(\frac{\bm{B} }{\rho} \right) \right|_\text{non-ideal} + \left.\frac{\rm d}{{\rm d}t} \left(\frac{\bm{B} }{\rho} \right) \right|_\text{art} \label{eq:ind}, \\
\rho \frac{\rm d}{{\rm d}t}\left( \frac{E}{\rho}\right) & = & -\nabla\cdot \bm{F} - \mbox{$\nabla \bm{v}${\bf :P}} + 4\pi \kappa \rho B_\text{P} - c \kappa \rho E, \label{eq:radiation} \\
\frac{{\rm d}u}{{\rm d}t} & = & -\frac{p}{\rho} \nabla\cdot\bm{v} - 4\pi \kappa B_\text{P} + c \kappa E\notag \\
                                             & + & \left.\frac{{\rm d} u}{\rm{d} t} \right|_\text{non-ideal} + \left.\frac{{\rm d} u}{\rm{d} t} \right|_\text{art}, \label{eq:matter} \\
\nabla^{2}\Phi & = & 4\pi G\rho, \label{eq:grav}
\end{eqnarray}
where ${\rm d}/{{\rm d}t} \equiv \partial/\partial t  + \bm{v}\cdot \nabla$ is the Lagrangian derivative,  $\rho$ is the density, ${\bm  v}$ is the velocity, $p$ is the gas pressure, ${\bm B}$ is the magnetic field, $\Phi$ is the gravitational potential, $B_\text{P}$ is the frequency-integrated Plank function, $E$ is the radiation energy density, $\mbox{\boldmath $F$}$ is the radiative flux, {\bf P} is the radiation pressure tensor, $c$ is the speed of light, $G$ is the gravitational constant, and $\mathbb{I}$ is the identity matrix.   The terms with subscript `non-ideal' are the contribution from the non-ideal MHD processes, and the terms with subscript `art' are artificial terms required for numerical stability.  We assume units for the magnetic field such that the Alfv{\'e}n speed is $v_{\rm A} = \vert B\vert/\sqrt{\rho}$ \cite{PriceMonaghan2004b}.

To evolve these equations, we use the three-dimensional smoothed particle hydrodynamics code \textsc{sphNG} that originated from W. Benz \cite{Benz1990}; over the past few decades, it has been substantially modified to include individual particle time-steps \cite{BateBonnellPrice1995}, variable smoothing lengths \citeparen{PriceMonaghan2004b,PriceMonaghan2007}, radiation \cite{BateKeto2015}, magnetohydrodynamics \cite{Price2012} and the non-ideal MHD processes \citeparen{WursterPriceAyliffe2014,WursterPriceBate2016,Wurster2016}.  We use the cubic spline smoothing kernel, thus each particle has approximately 58 neighbours. 

\subsection{Artificial algorithms}
Artificial algorithms are required for numerical stability, and include artificial viscosity, resistivity and conductivity that are applied to the momentum \eqref{eq:mom}, induction  \eqref{eq:ind} and internal energy \eqref{eq:matter} equations, respectively.  These terms are used to smooth discontinuities such that simulations remain stable.  However, applying too much artificial viscosity/resistivity/conductivity will smooth the domain too much such that the results are no longer physical; applying too little may not prevent instabilities from occurring, thus preventing reliable results.   For this paper, we will focus on the magnetic field and on artificial resistivity.

The discretised form of artificial resistivity we use (\!\!\cite{PriceMonaghan2004}, \cite{PriceMonaghan2005}, \cite{Phantom2018}) is
\begin{eqnarray}
\left.\frac{\text{d} }{\text{d} t}\left(\frac{B^i_a}{\rho_a}\right)\right|_\text{art}  = \frac{1}{\Omega_{a}\rho_{a}^{2}} \sum_{b} m_{b}  v_{\text{sig},ab}B^i_{ab}   \hat{r}^j_{ab} \nabla^j_aW_{ab}(h_{a}), \label{eq:artificialB}
\end{eqnarray}
where we sum over all particles $b$ within the kernel radius, $W_{ab}$ is the smoothing kernel, $m_b$ is the particle mass, $\Omega_a$ is a dimensionless correction term to account for a spatially variable smoothing length $h_a$,  $B^i_{ab} \equiv B^i_a - B^i_b$, and the signal velocity is  $v_{\text{sig},ab} = |\bm{v}_{ab} \times \hat{\bm{r}}_{ab} |$.  This artificial resistivity is second-order accurate away from shocks i.e. $\propto h^2$, thus, will decrease as $h$ decreases (i.e. as resolution increases).  For an investigation in to artificial resistivity prescriptions, see \cite{Wurster+2017}.

The remaining terms in the induction equation are also dependent on resolution (i.e. on $h$), but to a lesser degree; the ideal component of the induction equation (i.e. the first term of Equation \ref{eq:ind}) is given by
\begin{eqnarray}
\left.\frac{\text{d} }{\text{d} t}\left(\frac{B^i_a}{\rho_a}\right)\right|_\text{ideal} = -\frac{1}{\Omega_a \rho_a^2} \sum_b m_b   v^i_{ab} B^j_a \nabla^j_a W_{ab}\left(h_a\right).
\end{eqnarray}
Higher resolution models can calculate higher values of the magnetic field (if physically motivated) since the smoothing will occur over a smaller region and due to the decreased contribution from the artificial resistivity.  Thus, increased resolution will contribute to a more precise value. 

\section{Initial conditions}
\label{sec:ics}
Our initial conditions are similar to our previous studies \citeparen{BateTriccoPrice2014,WursterBatePrice2018sd,WursterBatePrice2018hd,WursterBatePrice2018ff}.  A sphere of radius $R=4\times~10^{16}$~cm and mass $M=1$~\Msun{} is placed in a  low-density box of edge length $l = 4R$ and a density contrast of 30:1.  The cloud has an initial rotational velocity of $\Omega = 1.77\times 10^{-13}$~rad~s$^{-1}$ and an initial sound speed of $c_\text{s,0} = 2.19\times 10^4$~cm~s$^{-1}$.  The entire domain is threaded with a uniform magnetic field strength of $B_0 = 163\mu$G which is parallel to the rotation axis.   The low-density box is used to provide boundary conditions at the edge of the cloud so that the entire cloud can be self-consistently modelled; periodic boundary conditions are used for the magnetohydrodynamic processes at the edge of the box, which is sufficiently far from the sphere such that the boundary conditions will have no effect on the results.  

The Hall effect, one of the non-ideal MHD processes, is dependent on the direction of the magnetic field \cite{WardleNg1999}, thus we present three models: an ideal MHD model (where the non-ideal terms are set to zero; named \iBzn{}), a non-ideal model where the magnetic field vector is aligned with the rotation vector (named \nBzp{}) and a non-ideal model where the vectors are anti-aligned (named \nBzn{}); collectively, the non-ideal models are \nBzpn{}.  For each model, we present resolutions of \ntf{}, \os{} and \ts{} particles in the sphere.  Model names without superscripts refer to all three models of the given initial conditions; names with a superscript refer to a particular model with the resolution stated in the superscript.

\section{Results}
\label{sec:results}

From previous star formation studies \citeparen{Tsukamoto+2015hall,Tsukamoto+2015oa,Tsukamoto+2017,WursterBatePrice2018sd,WursterBatePrice2018hd,WursterBatePrice2018ff}, the expected outcome of each model is known.  Model \iBzn{} is performed in strong, ideal magnetic fields and will succumb to the magnetic braking catastrophe \cite{AllenLiShu2003}, thus the collapse will be axi-symmetric  and no protostellar disc will form.  When including non-ideal MHD, the Hall effect will extract angular momentum from near the centre if the magnetic field is aligned with the rotation axis as in \nBzp{}, yielding an approximately axi-symmetric collapse and forming a small disc at late times; the Hall effect will contribute to the angular momentum near the centre if the magnetic field is anti-aligned with the rotation axis as in \nBzn{}, forming a large disc early during the collapse.  Although these are the expected outcomes, it has yet to be investigated how they will be affected by resolution.

Our analysis will be split into three components.  First, we will discuss the general evolution, which will follow the gravitational collapse of the gas from the initial density of \rhoapprox{-17} through the first hydrostatic core phase (also first core phase; \cite{Larson1969}) to the formation of the protostar at \rhoxapprox{-4} (\secref{sec:res:global}).  We will then investigate the gas and magnetic field structure near the end of the first core phase (\rhoapprox{-8}; \secref{sec:res:FHC}) and just prior to the birth of the protostar (\secref{sec:res:birth}).  

\subsection{Global evolution}
\label{sec:res:global}
\begin{figure}
\centering
\includegraphics[width=0.8\columnwidth]{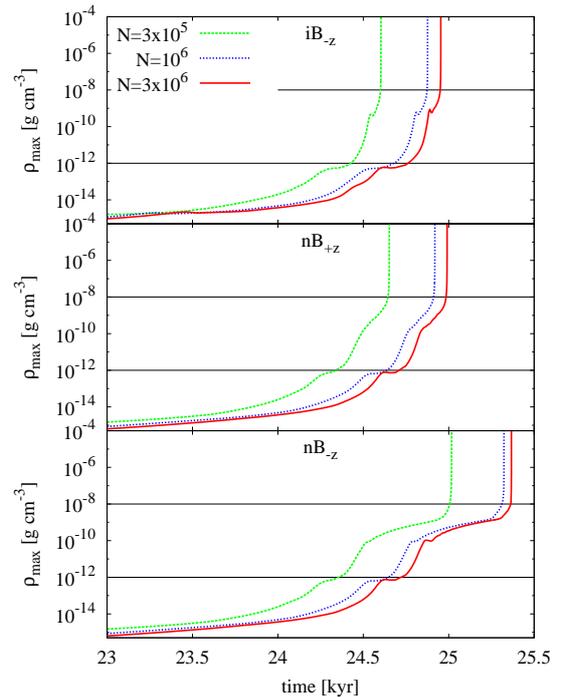}
\caption{Evolution of the maximum density during the gravitational collapse of a molecular cloud to form a protostar.  The lower and upper horizontal lines represent the beginning and end of the first hydrostatic core phase, respectively.  We define the birth of the protostar to occur at \rhoxeq{-4}.  At each resolution, there is a similar off-set in time between models reaching \rhoxapprox{-8}, and this off-set is decreasing with increasing resolution.  This suggests that the \ts{} models yield a prediction of the end of the first core phase that is only slightly early than the `true' time.}
\label{fig:rvt}
\end{figure}

As a molecular cloud gravitationally collapses, the central density increases, which is used to determine the phase of evolution.  Figure~\ref{fig:rvt} shows the maximum density as a function of time for each model.  
As expected from previous studies, all models at a given resolution initially collapse at similar rates, reaching the first core phase at \rhoxapprox{-12} at similar times.  The evolution then diverges, with \iBzn{} collapsing faster to the end of the first core phase at \rhoxapprox{-8} than \nBzp{} which collapses faster than \nBzn{}.   At each resolution, \iBzn{} reaches \rhoxapprox{-8} \sm30~yr before \nBzp{} and \sm400~yr before \nBzn{}.  Thus, the relative difference in collapse times between models is independent of resolution.  

For each model, the first core phase ends \sm290~yr in the \tf{} version prior to the \os{} version, which ends \sm60~yr prior to the \ts{} version.  Although these models have not converged on an end time, the trend suggests that the `true' end-time is only slightly later than that calculated in the \ts{} models, thus end time of the \ts{} models yield a reasonable prediction.

Given the varying collapse times, it is convenient to use maximum density as a proxy for time.  For the duration of this paper, we will compare the simulations at similar maximum densities, rather than at similar times.

As gas collapses in ideal MHD, the magnetic field is dragged into the centre with the gas, thus the strongest field strength is coincident with the maximum density \cite{BateTriccoPrice2014}.  Once non-ideal MHD processes are included, the collapse of the charged gas is slowed by the magnetic fields, while the neutral gas continues to collapse to the centre of the system. The charged gas is stalled in a torus around the central density peak, and the magnetic fields pile up here, creating a so-called `magnetic wall' \citeparen{LiMckee1996,TassisMouschovias2005b}. 

\begin{figure}
\centering
\includegraphics[width=0.8\columnwidth]{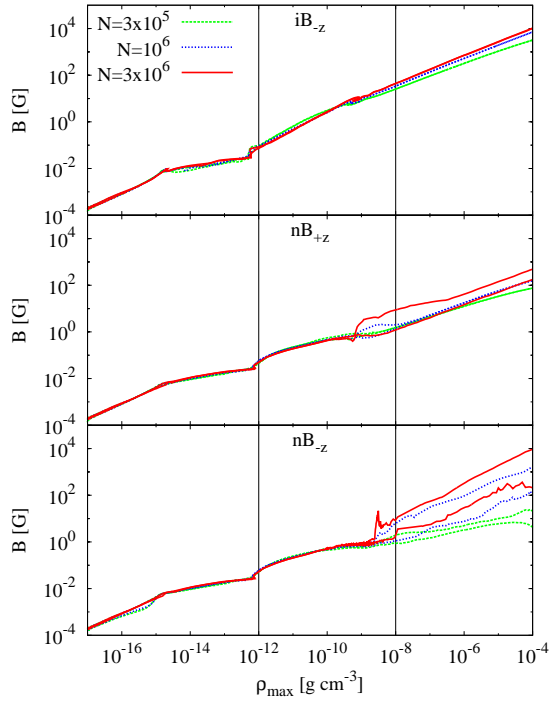}
\caption{Evolution of the maximum and central magnetic field strengths as a function of maximum density, which is a proxy for time.  Each model has two lines per resolution, where the upper line represents the maximum magnetic field strength $B_\text{max}$, and the lower line represents the central field strength $B_\text{cen}$ which is coincident with \rhox.  Where only one line exists, $B_\text{max} = B_\text{cen}$.  The vertical lines represents the beginning and end of the first hydrostatic core phase.  $B_\text{max} = B_\text{cen}$ in \iBzn{} (top panel) and for \rhoxls{-9} in the non-ideal models (bottom two panels).  The field strength is approximately independent of resolution for \iBzn{}, but strongly resolution-dependent for \nBzp{3e6} and \nBzn{} after \rhoxgs{-9}.}
\label{fig:bvr}
\end{figure}

Figure~\ref{fig:bvr} shows the maximum (\Bmax{}) and central (\Bcen{}; coincident with the maximum density \rhox) magnetic field strengths for each model; where only one line exists for a given model, then $B_\text{max} = B_\text{cen}$.

In ideal MHD, \iBzn{}, $B_\text{max} = B_\text{cen}$ as expected.  However, $B_\text{max}$ is approximately independent of resolution, and by \rhoxeq{-4}, $B_\text{max}$ differs by only a factor of \sm3 between \iBzn{3e5} and \iBzn{3e6}, and \sm1.4 between  \iBzn{1e6} and \iBzn{3e6}, suggesting that the value of \Bmax{} is converging.   Since \Bmax{} coincides with \rhox, then much of this decrease can be attributed to the artificial resistivity algorithms.  Despite a factor of \sm3 difference in magnetic field strengths, this difference is small in astrophysical terms.  Thus, we conclude, for ideal MHD, the maximum magnetic field strength up to the formation of a protostar is relatively insensitive to resolution.

When using non-ideal MHD, we expect a magnetic wall to form during the first core phase, thus \Bmax{} and \Bcen{} should diverge; this occurs at \rhoxapprox{-9} in \nBzpn{} (bottom two panels of Figure~\ref{fig:bvr}).  From this figure, it appears that a wall is formed and sustained, forms and dissipates, and never forms in the high, mid and low resolution \nBzp{} models, respectively; thus, by the formation of the protostar at \rhoxapprox{-4}, $B_\text{max} \ne B_\text{cen}$ in \nBzp{3e6} only.  \Bcen{} is approximately the same in the high and mid resolution model, which is approximately \sm2 times higher than the low resolution model.  Although $B_\text{cen}$ is similar at the birth of the protostar, different conclusions are reached regarding \Bmax{} and the magnetic wall based upon the initial resolution.

At each resolution in \nBzn{} (bottom panel of Figure~\ref{fig:bvr}), there is a clear distinction between $B_\text{max}$ and $B_\text{cen}$ for \rhoxgs{-9}.  In this model, $B_\text{max}$ ($B_\text{cen}$) differs by a factor of \sm5000 (\sm50) amongst the three resolutions.  As the resolution decreases, the ratio $B_\text{max}/B_\text{cen}$ also decreases.  Thus, this model clearly has a strong dependence on resolution; the current results suggest that we cannot predict to what values \Bmax{} and \Bcen{} will converge.

These non-ideal MHD plots suggest that resolution is very important in reaching the correct conclusions, and that the convergence limit has not yet been reached.  This resolution-dependence is somewhat counter-intuitive since the non-ideal models include additional resolution-insensitive resistivity processes that should make the results less susceptible to resolution effects.    However, we need to be cautious with the above analysis since it involves only one or two values at each time or density.  In each of the next two sections, we will focus on the gas structure at a single time, which will complement the above time-sequence analysis.

\subsection{First hydrostatic core}
\label{sec:res:FHC}
\begin{figure}
\centering
\includegraphics[width=\columnwidth]{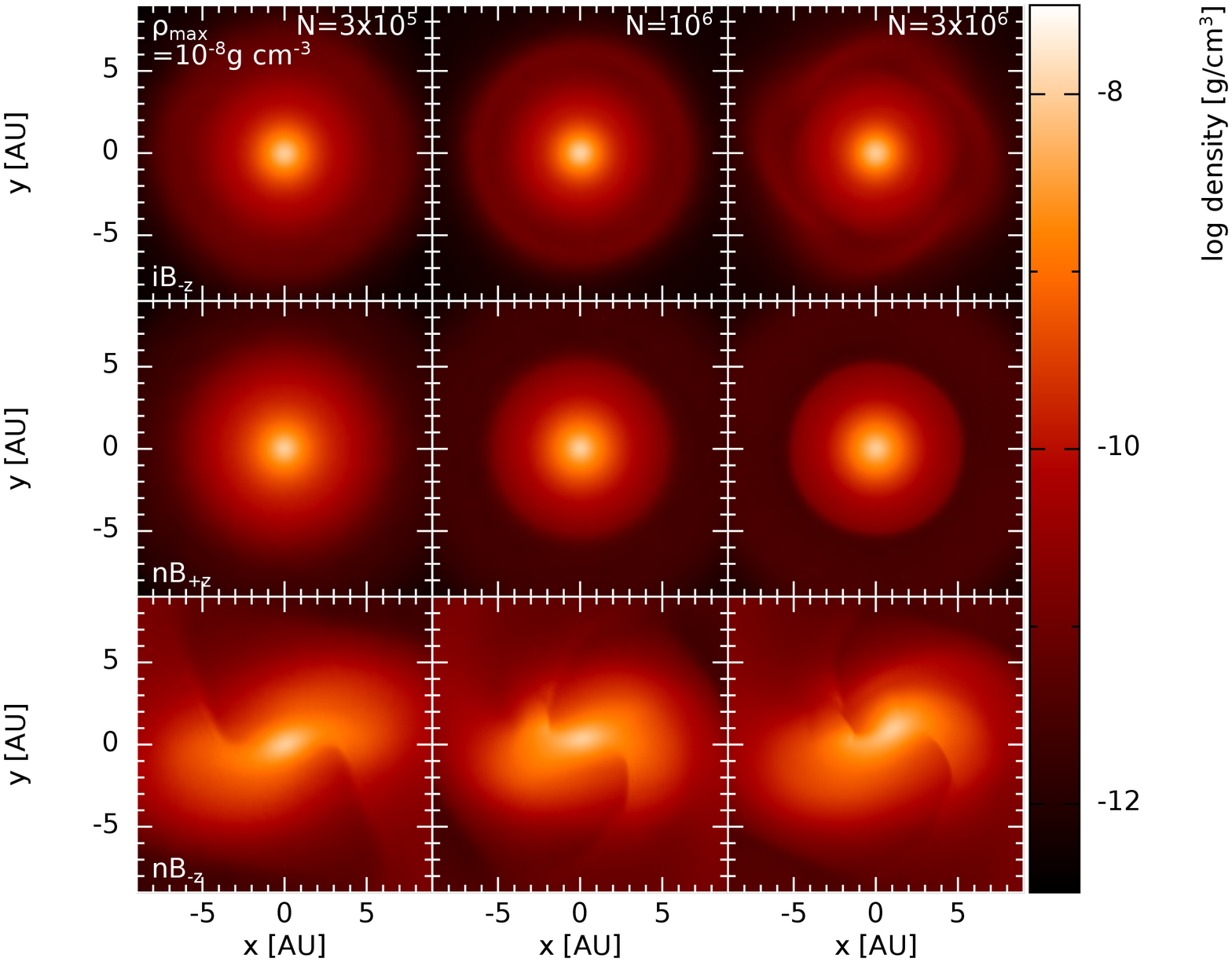}
\includegraphics[width=\columnwidth]{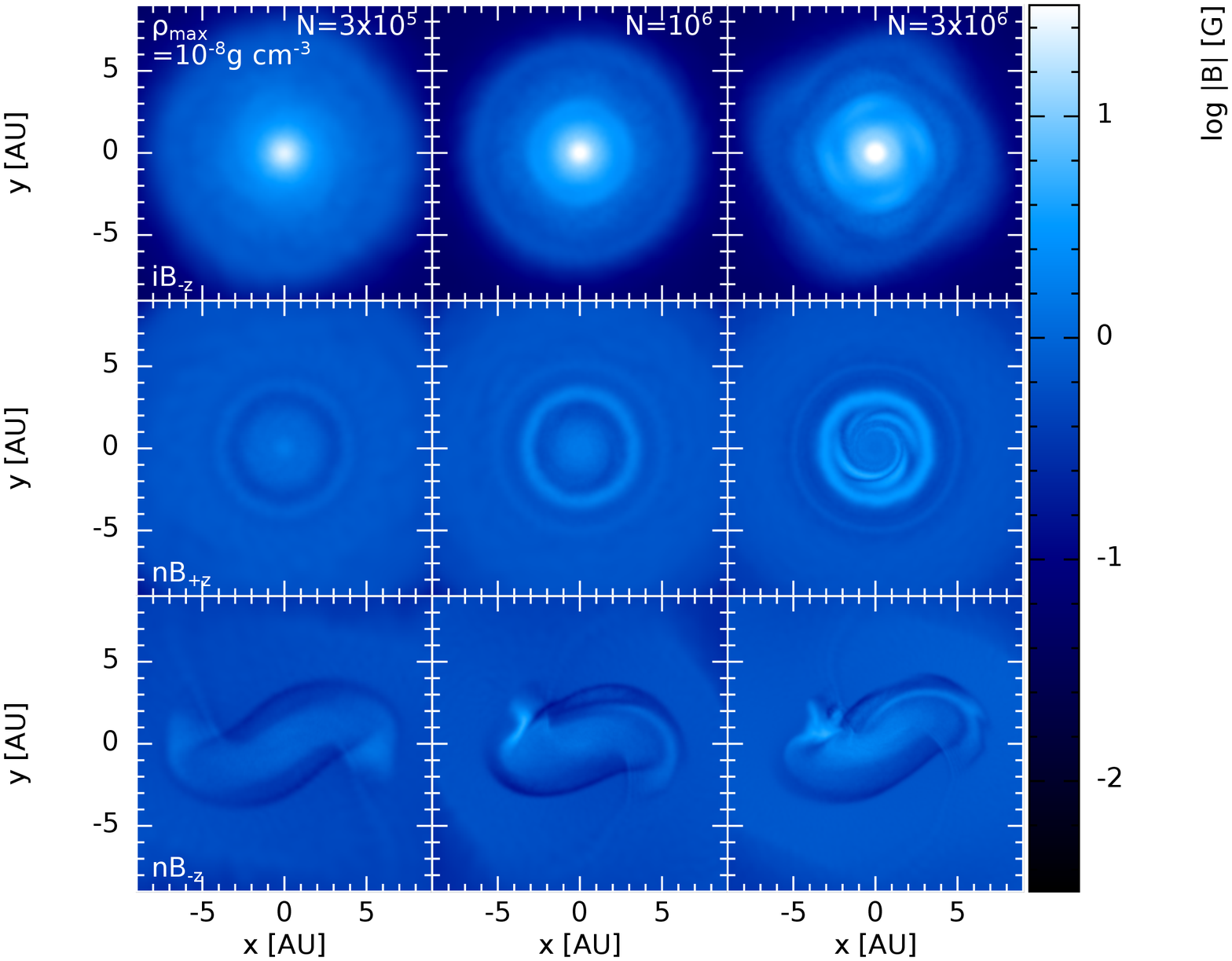}
\caption{Gas density (top panel) and magnetic field strength (bottom panel) at the end of the first hydrostatic core phase at \rhoxeq{-8}.  The frames sizes are selected to show the structure of the magnetic wall (middle row, bottom panel) and of the bar (bottom row, both panels).  The density profiles have minimal resolution dependence.  The non-ideal magnetic field geometries are strongly dependent on resolution, in particular the strength and structure of the magnetic wall (model \nBzp{}) and the location and structure of the maximum magnetic field strength (model \nBzn{}).}
\label{fig:FHC}
\end{figure}

The first hydrostatic core phase is $10^{-12} \lesssim \rho_\text{max}/($\gpercc$) \lesssim 10^{-8}$ \cite{Larson1969}.  From Figure~\ref{fig:bvr}, the models appear similar the beginning of this phase, thus we will analyse the density and magnetic field at the end of this phase once \Bmax{} and \Bcen{} become distinct.  Figure~\ref{fig:FHC} shows the gas density and magnetic field strength through the mid-plane near the end of the first core phase at \rhoxapprox{-8}.

For the inner $r\lesssim 10$~au, the density profiles of \iBzn{} and \nBzp{} are approximately axi-symmetric and approximately independent of resolution.  The additional physical processes in \nBzn{} have permitted a gravitationally unstable disc to form at \rhoxtwoapprox{5}{-9}; the disc has a radius of $r\approx25$~au, thus only the central bar is shown in Figure~\ref{fig:FHC}.  The bar structure is similar at each resolution, although the current rotation angle is different simply due to the different lengths of time it has taken to reach the current density (recall Section~\ref{sec:res:global}); note that the rotational velocities are similar at each resolution.   Once the rotation angle is taken into account, then near the end of the first core phase, the density structure of the inner $r\lesssim 10$~au is approximately independent of resolution.

For all three models for $r \gtrsim10$~au, the density profiles become more broad for increasing resolution; specifically, the higher resolution models can capture more detail further from the core.  In \iBzn{} and \nBzp{}, this likely has only moderate consequence to any analyses since studies such as this typically focus on the first core itself (i.e. $r\lesssim 10$~au); however, studies such as \nBzn{} typically focus on the disc thus require good resolution out to several tens of au, thus resolution at larger radii must also be carefully considered.

It is clear from Figure~\ref{fig:FHC} that both the magnetic strength and structure are affected by the both the physical processes and resolution.  The magnetic field structure of \iBzn{} is least dependent on resolution. However, the non-ideal MHD models \nBzpn{} are strongly dependent on resolution, despite their additional resistive processes being physically motivated.  In each of the \nBzp{} models (middle row), there is a torus of gas with a strong magnetic field that comprises of the magnetic wall, whose maximum extent is $r \sim4$~au.  In \nBzp{3e5}, the wall has a slightly higher magnetic field strength than the surrounding gas, but is still lower than \Bcen{} = \Bmax{} (recall Figure~\ref{fig:bvr}).  Increasing the resolution to \os{}, the magnetic wall becomes well-defined.  This is a small wall of width d$r\sim1.5$~au, but its maximum magnetic field strength is only slightly higher than then central strength.  From this snapshot, the processes that are causing the magnetic wall are just defined at this resolution.  However, given the weakness of the wall, their full effect cannot be fully understood.

The magnetic wall is very well-defined in our highest resolution model, \nBzp{3e6}.  As in \nBzp{1e6}, the primary wall has width d$r\sim1.5$~au and the outermost extent is $r\sim4$~au.  In the high resolution wall, the magnetic field strength is \sm5 times stronger than \Bcen{}.  More importantly,  spiral structures are visible interior to $r \lesssim 4$~au, and these spirals have field strength \sm10 times higher than \Bcen{}.  These spirals are caused by the rotating, highly magnetised gas on the interior of the torus losing angular momentum, detaching from the torus and migrating towards the centre.  This spiral structure is absent in the density plot (top panel) since the majority of the gas is neutral and acting mostly independently of the magnetic field.  Thus, these spirals are created by the small fraction of charged gas that drags the magnetic field into the centre, which cannot be distinguished in the density plots.

Thus, for \nBzp{}, there is no discernible magnetic wall for \nBzp{3e5}, a defined wall for \nBzp{1e6}, and a very well-defined wall for \nBzp{3e6} where the gas can be seen detaching from the wall and spiralling towards the centre.  Although \nBzp{1e6} yields the wall, these results suggest that the higher resolution of \ts{} is required to resolve the behaviour of the charged gas in and leaving the torus.  Since we did not perform a higher resolution model, we cannot confidently say that even \ts{} is high enough to properly resolve the formation and evolution of the magnetic wall.

The gravitationally unstable disc that forms during the first core phase when \rhoxtwoapprox{5}{-9}  in \nBzn{} complicates the evolution of the magnetic field; the formation of this disc coincides with the divergence of \Bmax{} and \Bcen{}.  Contrary to what was discussed in Section~\ref{sec:res:global} and predicted by Figure~\ref{fig:bvr}, there is no magnetic wall.  In this model, the magnetic field tries to anchor the gas, thus a bar of charged gas slightly lags behind the primary, mostly neutral bar; this is unseen in the density plot of Figure~\ref{fig:FHC} since the density of charged gas is much less than that of neutral gas.  Thus, the motion of the charged and neutral gas caused by the bars prevent the formation of a magnetic wall; rather, the most magnetised gas initially piles up near the edge of the bar.  Thus, the divergence of \Bmax{} and \Bcen{} in this model is due to the rotation of the bar rather than a magnetic wall.

In \nBzn{3e5}, the bar has nearly a uniform magnetic field strength, but in \nBzn{1e6} and \nBzn{3e6}, a strongly magnetised clump of gas has collapsed to the west side of the bar.  Although symmetric results would have been expected given the initial conditions, even slight asymmetries caused by numerical processes or rounding could cause the discrepancy that has lead to the asymmetric bar.  Independent of the cause, the asymmetry has formed in similar locations in both models, with the structure being better well-defined and having a stronger magnetic field strength in \nBzn{3e6}.

In very general terms, the three \nBzn{} models are qualitatively similar, yielding a \sm25~au disc during the first core phase.  The bar structure is similar amongst them, however, the asymmetry only appears at the higher two resolutions, suggesting a minimum resolution  of \os{}.

\subsection{Stellar birth}
\label{sec:res:birth}
\begin{figure}
\centering
\includegraphics[width=\columnwidth]{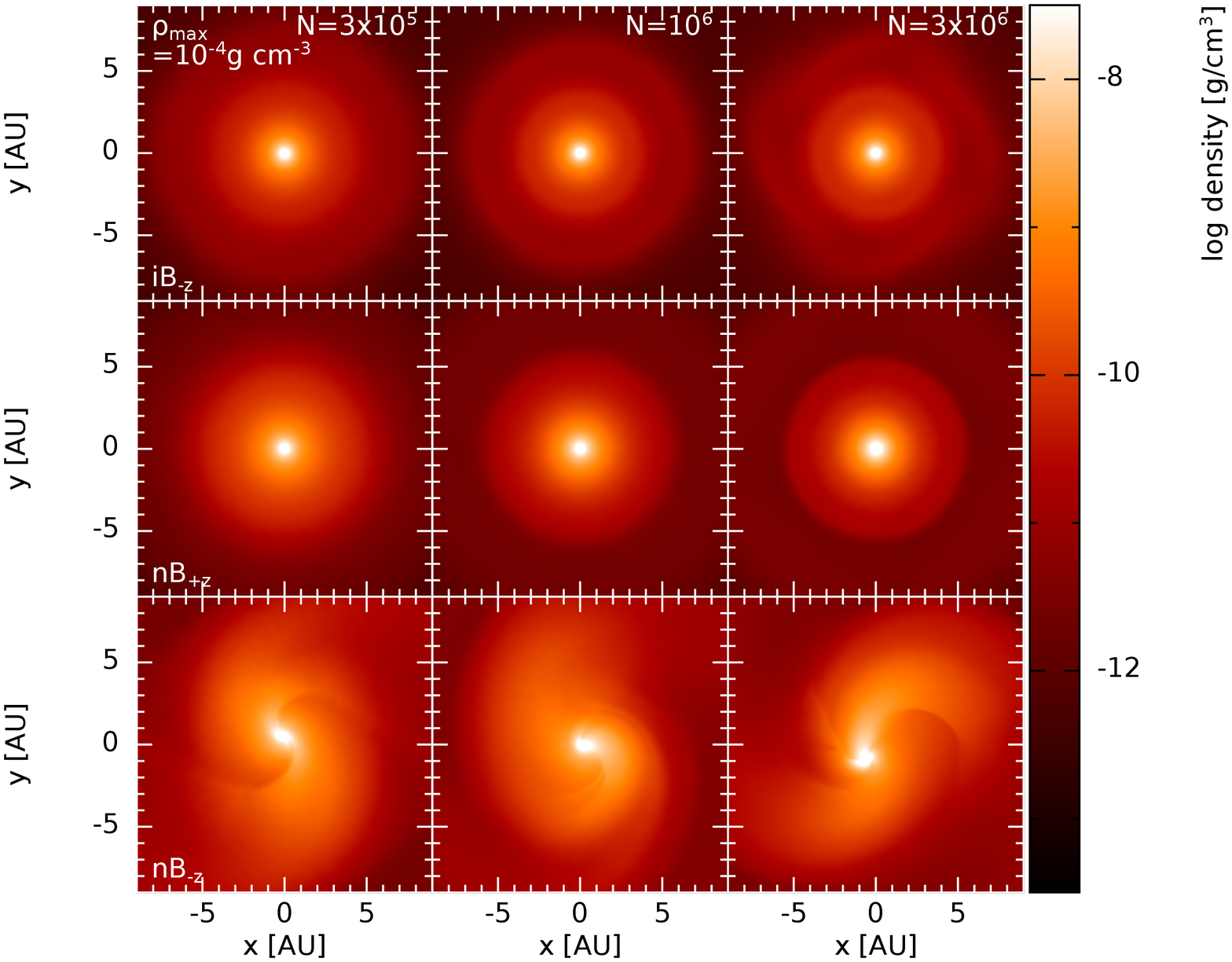}
\includegraphics[width=\columnwidth]{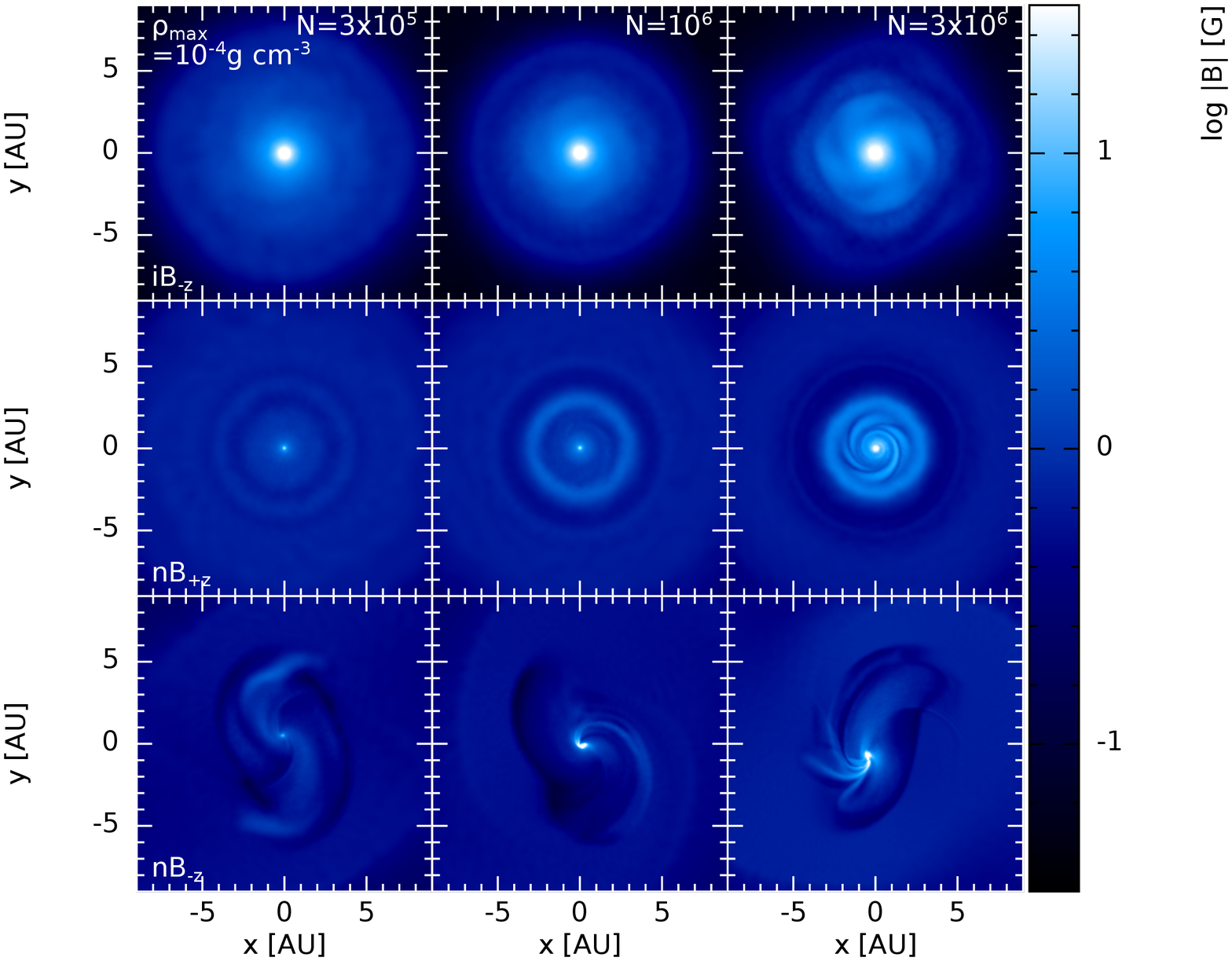}
\caption{Gas density (top panel) and magnetic field strength (bottom panel) at the birth of the protostar at \rhoxeq{-4}.  In \nBzp{} (middle row, bottom panel), the magnetic wall persists, although its magnetic field strength is weaker than the central field strength.  The bar has begun to collapse in \nBzn{} (bottom row, both panels), beginning to form a spherical core with a strong magnetic field strength.}
\label{fig:sb}
\end{figure}

Between the end of the first core phase at \rhoxapprox{-8} and the birth of the protostar at \rhoxapprox{-4}, only a few years pass (recall Figure~\ref{fig:rvt}).  However, there is noticeable evolution during this time, specifically in the centre of the cloud and in the non-ideal models.  Figure~\ref{fig:sb} shows the gas density and magnetic field strength for each model at \rhoxeq{-4}.

Similar to the end of the first core phase, the density profiles of \iBzn{} and \nBzp{} are approximately independent of resolution.  During this time, the bar in \nBzn{} collapses and this collapse is resolution-dependent: it stays approximately symmetric for \nBzn{3e5}, collapses asymmetrically to near the centre in \nBzn{1e6}, and collapses to the west end of the bar in \nBzn{3e6}.  Thus, for the first time in this study, we see noticeable resolution effects on the structure of the density.  

The magnetic field structure of \iBzn{} is similar at all resolutions, but the field strength decreases with resolution, especially in the inner \sm0.01~au.  This decrease is due to a combination of the lower resolution smoothing out the field strength, and the higher artificial resistivity dissipating more of the magnetic field.

Model \nBzp{} has retained its magnetic wall (and spiral structure interior to it for \nBzp{3e6}).  The magnetic wall remains well defined in the two higher resolution cases, and in all cases, its field strength is stronger than that in the surrounding gas.   Thus, independent of resolution, we can conclude that the magnetic wall is a persistent feature.  In \nBzp{3e5} and \nBzp{1e6}, the \Bmax{} = \Bcen{}, again with a lower strength in the former model.  In \nBzp{3e6}, the maximum field strength is \sm3 times stronger than the central strength, but the maximum strength is only slightly mis-coincident with the maximum density.  Thus, the expected location of \Bmax{} at this time cannot be determined; although it is coincident with the maximum density in two of the models, it is separate from the maximum density in the highest resolution model.  Thus even higher resolution is required to reach a robust conclusion on the expected location of \Bmax{}.

As with the density evolution, the magnetic field structure of \nBzn{} is highly dependent on resolution.  None of these models had a well-defined magnetic wall, nor has one formed after the end of the first core phase.  In all cases, the maximum magnetic field strength resides 0.01-0.1~au from the maximum density, and the off-set decreases with resolution.  Thus, although the gas is collapsing along the bar, the magnetic field is not necessarily tracing this collapse.   Although the off-set decreases with increasing resolution, there is likely a minimum, non-zero off-set which will be reached at the convergence limit since there is an initial off-set during the first core phase and non-zero off-sets are possible given the ideal MHD model, \iBzn{}.  The values of \Bmax{} and \Bcen{} increase with increasing resolution, as does the factor between them, indicating that artificial resistivity is less important with increasing resolution.  Again, we cautiously conclude that \Bmax{} $\ne$ \Bcen{} is a real effect, but we cannot comment on the values that these terms should take, nor the expected location of \Bmax{}.  
	
\section{Discussion and Summary}
\label{sec:conc}
\subsection{Computational expense}
\begin{figure}
\centering
\includegraphics[width=\columnwidth]{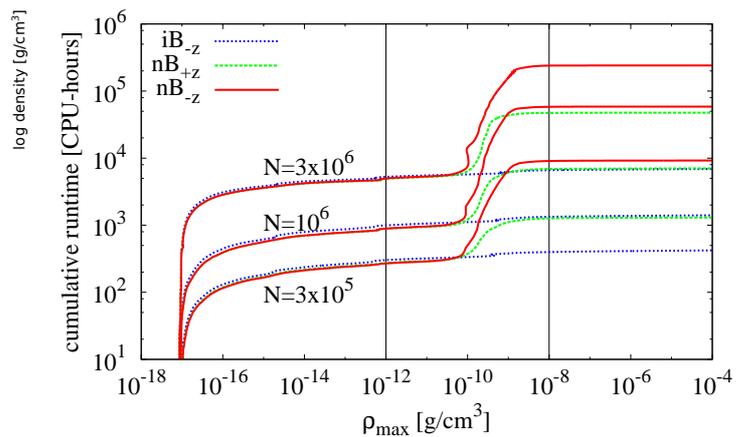}
\caption{The cumulative CPU time used for each model. The vertical lines are at the beginning and end of the first hydrostatic core phase.  Independent of the non-ideal processes, all models at a given resolution take a similar length of computational time to reach \rhoxapprox{-10}, after which \nBzn{} and \nBzp{} require considerable more computational resources to evolve through the first core phase.}
\label{fig:cpu}
\end{figure}

We have repeatedly stated that higher resolution simulations are required to reach the convergence limit and robust conclusions.  However, higher resolution simulations necessarily take more resources, which will eventually become a prohibiting factor.  Figure~\ref{fig:cpu} shows the runtime for each model at each resolution.

For \iBzn{}, the cumulative CPU time increases smoothly, and \iBzn{3e6} takes only 7000~CPU-hours to reach the formation of the protostar.  Thus, if we were to increase the resolution by a factor of 10 to $N=3\times10^7$, we can extrapolate the runtime to be $\sim5\times10^5$~CPU-hours.  However, the non-ideal models have greater physical motivation, and Figure~\ref{fig:cpu} shows that evolving the first core phase is computationally demanding in \nBzpn{}.  Even at \nts{} it takes \nBzp{3e6} and \nBzn{3e6} $4.9\times 10^4$ and $2.4\times 10^5$~CPU-hours to reach \rhoxeq{-4}, respectively.  Increasing resolution by a factor of 10 would require nearly $10^7$~CPU-hours for \nBzn{3e7}.  Thus, convergence studies on star formation will be computationally expensive, thus comparisons such as those presented here are required to determine the acceptable trade-off between accuracy and computational expense.

Although already expensive, there is great interest in modelling the early evolution of the protostar itself \citeparen{BateTriccoPrice2014,WursterBatePrice2018hd,WursterBatePrice2018ff}, which requires reaching densities of at least \rhoxapprox{-1}.  However, the computational time to model the evolution between $10^{-2} \lesssim \rho_\text{max}/($\gpercc$) \lesssim 10^{-1}$ increases considerable for all models \cite{WursterBatePrice2018ff}, making the search for the convergence limit in this scientifically interesting regime even more prohibitive. 

Counter-intuitively, there is less resolution convergence in the non-ideal models than in the ideal model.  The non-ideal processes introduce a restrictive time-stepping criteria $\propto~h^2$ \citeparen{Maclow+1995,ChoiKimWiita2009,WursterPriceAyliffe2014}  which is always satisfied by construction and accounts for the steep increase in runtime during the first core phase; for reference, the Courant time-step is longer and is $\propto~h$.  These processes also introduce a new characteristic length scale \citeparen{WursterPriceAyliffe2014,MarchandCommerconChabrier2018}, which we have determined is resolved in the magnetic wall and the bar.  Given the lack of convergence, it is possible that these new length scales are not restrictive enough to properly resolve the non-ideal MHD processes, and that new characteristic scales need to be determined.  However, this will likely require finding the convergence limit, which may be prohibitively expensive.

\subsection{Summary}
We ran three star formation models each at three different resolutions.  We performed one ideal MHD simulation (\iBzn{}), and two non-ideal MHD simulations (\nBzp{} and \nBzn{}) using two different initial alignments of the magnetic field to test the extreme effects of the non-ideal process of the Hall effect.   In all models, the higher resolution simulations collapsed later than lower resolution simulations, but at a given resolution, the relative collapse time between models was consistent.  

The general structure of the ideal MHD models was similar at each resolution, but with decreasing maximum magnetic field strength for decreasing resolution.  Both the lower resolution and the greater contribution from artificial resistivity contributed to this decrease.  

The magnetic field structure of the non-ideal MHD models was resolution-dependent.  At the end of the first hydrostatic core phase, \nBzp{} formed a magnetic wall which persisted to the formation of the protostar (although only very weakly in \nBzp{3e5}).  The magnetic field strength of the wall and the ability to model the gas being accreted from the inner edge of the wall was strongly dependent on resolution.  As the gas collapsed to form the protostar, the maximum magnetic field strength remained off-set from the maximum density only in the highest resolution simulation and the maximum strength was interior to the magnetic wall.  

A magnetic wall did not form prior to the birth of the protostar in \nBzn{}.  The maximum magnetic field strength was never coincident with the maximum density after the formation of the bar, and the maximum and central magnetic field strengths and the ratio between them was dependent on resolution.  The central bar in the disc collapsed after the first core phase, and the nature of the collapse was also sensitive to resolution.

Despite additional physical processes in the non-ideal models, these models were more dependent on resolution than the ideal MHD model.  Thus, these new processes necessarily introduced new time and length scales that must be resolved; although these are resolved in all our non-ideal models, they may not be restrictive enough to fully resolve these processes.   

The accuracy obtained from higher resolution is always a trade-off with computational expense, and some very high resolution simulations are infeasible to run.  In any resolution study, the convergence limit would ideally be reached, but this limit has often proved elusive in astrophysical studies.  Therefore, we must always be cautious of the results, and perform simulations as feasible or as required to examine the impact of resolution on all conclusions.

\section*{Acknowledgment}
JW and MRB acknowledge support from the European Research Council under the European Community's Seventh Framework Programme (FP7/2007- 2013 grant agreement no. 339248).  Calculations and analyses for this paper were performed on the University of Exeter Supercomputer, Isca, which is part of the University of Exeter High-Performance Computing (HPC) facility, and on the DiRAC Data Intensive service at Leicester, operated by the University of Leicester IT Services, which forms part of the STFC DiRAC HPC Facility (www.dirac.ac.uk). The equipment was funded by BEIS capital funding via STFC capital grants ST/K000373/1 and ST/R002363/1 and STFC DiRAC Operations grant ST/R001014/1. DiRAC is part of the National e-Infrastructure.  The research data supporting this publication are openly available from the University of Exeter's institutional repository at https://doi.org/10.24378/exe.607.  Several figures were made using \textsc{splash} \cite{Price2007}.  


\bibliographystyle{IEEEtran.bst}
\bibliography{SPHERICXIV_Wurster_Bate}

\begin{thebibliography}{10}
\providecommand{\url}[1]{#1}
\csname url@samestyle\endcsname
\providecommand{\newblock}{\relax}
\providecommand{\bibinfo}[2]{#2}
\providecommand{\BIBentrySTDinterwordspacing}{\spaceskip=0pt\relax}
\providecommand{\BIBentryALTinterwordstretchfactor}{4}
\providecommand{\BIBentryALTinterwordspacing}{\spaceskip=\fontdimen2\font plus
\BIBentryALTinterwordstretchfactor\fontdimen3\font minus
  \fontdimen4\font\relax}
\providecommand{\BIBforeignlanguage}[2]{{%
\expandafter\ifx\csname l@#1\endcsname\relax
\typeout{** WARNING: IEEEtran.bst: No hyphenation pattern has been}%
\typeout{** loaded for the language `#1'. Using the pattern for}%
\typeout{** the default language instead.}%
\else
\language=\csname l@#1\endcsname
\fi
#2}}
\providecommand{\BIBdecl}{\relax}
\BIBdecl

\bibitem{Wurster+2017}
J.~{Wurster}, M.~R. {Bate}, D.~J. {Price}, and T.~S. {Tricco}, ``{Investigating
  prescriptions for artificial resistivity in smoothed particle
  magnetohydrodynamics},'' \emph{ArXiv e-prints}, Jun. 2017.

\bibitem{BateBurkert1997}
M.~R. {Bate} and A.~{Burkert}, ``{Resolution requirements for smoothed particle
  hydrodynamics calculations with self-gravity},'' \emph{\mnras}, vol. 288, pp.
  1060--1072, Jul. 1997.

\bibitem{Nelson2006}
A.~F. {Nelson}, ``{Numerical requirements for simulations of self-gravitating
  and non-self-gravitating discs},'' \emph{\mnras}, vol. 373, pp. 1039--1073,
  Dec. 2006.

\bibitem{MeruBate2011criteria}
F.~{Meru} and M.~R. {Bate}, ``{On the fragmentation criteria of
  self-gravitating protoplanetary discs},'' \emph{\mnras}, vol. 410, pp.
  559--572, Jan. 2011.

\bibitem{MeruBate2011converge}
------, ``{Non-convergence of the critical cooling time-scale for fragmentation
  of self-gravitating discs},'' \emph{\mnras}, vol. 411, pp. L1--L5, Feb. 2011.

\bibitem{MeruBate2012}
------, ``{On the convergence of the critical cooling time-scale for the
  fragmentation of self-gravitating discs},'' \emph{\mnras}, vol. 427, pp.
  2022--2046, Dec. 2012.

\bibitem{Meyer+2018}
D.~M.-A. {Meyer}, R.~{Kuiper}, W.~{Kley}, K.~G. {Johnston}, and E.~{Vorobyov},
  ``{Forming spectroscopic massive protobinaries by disc fragmentation},''
  \emph{\mnras}, vol. 473, pp. 3615--3637, Jan. 2018.

\bibitem{Price2012turb}
D.~J. {Price}, ``{Resolving high Reynolds numbers in smoothed particle
  hydrodynamics simulations of subsonic turbulence},'' \emph{\mnras}, vol. 420,
  pp. L33--L37, Feb. 2012.

\bibitem{TriccoPriceFederrath2016}
T.~S. {Tricco}, D.~J. {Price}, and C.~{Federrath}, ``{A comparison between grid
  and particle methods on the small-scale dynamo in magnetized supersonic
  turbulence},'' \emph{\mnras}, vol. 461, pp. 1260--1275, Sep. 2016.

\bibitem{BoothClarke2019}
R.~A. {Booth} and C.~J. {Clarke}, ``{Characterizing gravito-turbulence in 3D:
  turbulent properties and stability against fragmentation},'' \emph{\mnras},
  vol. 483, pp. 3718--3729, Mar. 2019.

\bibitem{BateTriccoPrice2014}
M.~R. {Bate}, T.~S. {Tricco}, and D.~J. {Price}, ``{Collapse of a molecular
  cloud core to stellar densities: stellar-core and outflow formation in
  radiation magnetohydrodynamic simulations},'' \emph{\mnras}, vol. 437, pp.
  77--95, Jan. 2014.

\bibitem{WursterPriceBate2016}
J.~{Wurster}, D.~J. {Price}, and M.~R. {Bate}, ``{Can non-ideal
  magnetohydrodynamics solve the magnetic braking catastrophe?}''
  \emph{\mnras}, vol. 457, pp. 1037--1061, Mar. 2016.

\bibitem{WursterBatePrice2018ff}
J.~{Wurster}, M.~R. {Bate}, and D.~J. {Price}, ``{On the origin of magnetic
  fields in stars},'' \emph{\mnras}, vol. 481, pp. 2450--2457, Dec. 2018.

\bibitem{BrioWu1988}
M.~{Brio} and C.~C. {Wu}, ``{An upwind differencing scheme for the equations of
  ideal magnetohydrodynamics},'' \emph{J. Comp. Phys.}, vol.~75, pp. 400--422,
  Apr. 1988.

\bibitem{RyuJones1995}
D.~{Ryu} and T.~W. {Jones}, ``{Numerical magetohydrodynamics in astrophysics:
  Algorithm and tests for one-dimensional flow`},'' \emph{\apj}, vol. 442, pp.
  228--258, Mar. 1995.

\bibitem{OrszagTang1979}
S.~A. {Orszag} and C.-M. {Tang}, ``{Small-scale structure of two-dimensional
  magnetohydrodynamic turbulence},'' \emph{J. Fluid Mech.}, vol.~90, pp.
  129--143, Jan. 1979.

\bibitem{MestelSpitzer1956}
L.~{Mestel} and L.~{Spitzer}, Jr., ``{Star formation in magnetic dust
  clouds},'' \emph{\mnras}, vol. 116, p. 503, 1956.

\bibitem{HeilesCrutcher2005}
C.~{Heiles} and R.~{Crutcher}, ``{Magnetic Fields in Diffuse HI and Molecular
  Clouds},'' in \emph{Cosmic Magnetic Fields}, ser. Lecture Notes in Physics,
  Berlin Springer Verlag, R.~{Wielebinski} and R.~{Beck}, Eds., vol. 664, 2005,
  p. 137.

\bibitem{BateKeto2015}
M.~R. {Bate} and E.~R. {Keto}, ``{Combining radiative transfer and diffuse
  interstellar medium physics to model star formation},'' \emph{\mnras}, vol.
  449, pp. 2643--2667, May 2015.

\bibitem{Price2012}
D.~J. {Price}, ``{Smoothed particle hydrodynamics and magnetohydrodynamics},''
  \emph{Journal of Computational Physics}, vol. 231, pp. 759--794, Feb. 2012.

\bibitem{WardleNg1999}
M.~{Wardle} and C.~{Ng}, ``{The conductivity of dense molecular gas},''
  \emph{\mnras}, vol. 303, pp. 239--246, Feb. 1999.

\bibitem{BraidingWardle2012accretion}
C.~R. {Braiding} and M.~{Wardle}, ``{The Hall effect in accretion flows},''
  \emph{\mnras}, vol. 427, pp. 3188--3195, Dec. 2012.

\bibitem{BraidingWardle2012sf}
------, ``{ The Hall effect in star formation},'' \emph{\mnras}, vol. 422, pp.
  261--281, May 2012.

\bibitem{Phantom2018}
D.~J. {Price}, J.~{Wurster}, T.~S. {Tricco}, C.~{Nixon}, S.~{Toupin},
  A.~{Pettitt}, C.~{Chan}, D.~{Mentiplay}, G.~{Laibe}, S.~{Glover}, C.~{Dobbs},
  R.~{Nealon}, D.~{Liptai}, H.~{Worpel}, C.~{Bonnerot}, G.~{Dipierro},
  G.~{Ballabio}, E.~{Ragusa}, C.~{Federrath}, R.~{Iaconi}, T.~{Reichardt},
  D.~{Forgan}, M.~{Hutchison}, T.~{Constantino}, B.~{Ayliffe}, K.~{Hirsh}, and
  G.~{Lodato}, ``{Phantom: A Smoothed Particle Hydrodynamics and
  Magnetohydrodynamics Code for Astrophysics},'' \emph{\pasa}, vol.~35, p.
  e031, Sep. 2018.

\bibitem{PriceMonaghan2004b}
D.~J. {Price} and J.~J. {Monaghan}, ``{Smoothed Particle Magnetohydrodynamics -
  II. Variational principles and variable smoothing-length terms},''
  \emph{\mnras}, vol. 348, pp. 139--152, Feb. 2004.

\bibitem{Benz1990}
W.~{Benz}, ``{Smooth Particle Hydrodynamics - a Review},'' in \emph{Numerical
  Modelling of Nonlinear Stellar Pulsations Problems and Prospects}, J.~R.
  {Buchler}, Ed., 1990, p. 269.

\bibitem{BateBonnellPrice1995}
M.~R. {Bate}, I.~A. {Bonnell}, and N.~M. {Price}, ``{Modelling accretion in
  protobinary systems},'' \emph{\mnras}, vol. 277, pp. 362--376, Nov. 1995.

\bibitem{PriceMonaghan2007}
D.~J. {Price} and J.~J. {Monaghan}, ``{An energy-conserving formalism for
  adaptive gravitational force softening in smoothed particle hydrodynamics and
  N-body codes},'' \emph{\mnras}, vol. 374, pp. 1347--1358, Feb. 2007.

\bibitem{WursterPriceAyliffe2014}
J.~{Wurster}, D.~J. {Price}, and B.~{Ayliffe}, ``{Ambipolar diffusion in
  smoothed particle magnetohydrodynamics},'' \emph{\mnras}, vol. 444, pp.
  1104--1112, Oct. 2014.

\bibitem{Wurster2016}
J.~{Wurster}, ``{NICIL: A Stand Alone Library to Self-Consistently Calculate
  Non-Ideal Magnetohydrodynamic Coefficients in Molecular Cloud Cores},''
  \emph{\pasa}, vol.~33, p. e041, Sep. 2016.

\bibitem{PriceMonaghan2004}
D.~J. {Price} and J.~J. {Monaghan}, ``{Smoothed Particle Magnetohydrodynamics -
  I. Algorithm and tests in one dimension},'' \emph{\mnras}, vol. 348, pp.
  123--138, Feb. 2004.

\bibitem{PriceMonaghan2005}
------, ``{Smoothed Particle Magnetohydrodynamics - III. Multidimensional tests
  and the $\nabla \cdot B = 0$ constraint},'' \emph{\mnras}, vol. 364, pp.
  384--406, Dec. 2005.

\bibitem{WursterBatePrice2018sd}
J.~{Wurster}, M.~R. {Bate}, and D.~J. {Price}, ``{ The collapse of a molecular
  cloud core to stellar densities using radiation non-ideal
  magnetohydrodynamics},'' \emph{\mnras}, vol. 475, pp. 1859--1880, Apr. 2018.

\bibitem{WursterBatePrice2018hd}
------, ``{Hall effect-driven formation of gravitationally unstable discs in
  magnetized molecular cloud cores},'' \emph{\mnras}, vol. 480, pp. 4434--4442,
  Nov. 2018.

\bibitem{Tsukamoto+2015hall}
Y.~{Tsukamoto}, K.~{Iwasaki}, S.~{Okuzumi}, M.~N. {Machida}, and S.~{Inutsuka},
  ``{Bimodality of Circumstellar Disk Evolution Induced by the Hall Current},''
  \emph{\apjl}, vol. 810, p. L26, Sep. 2015.

\bibitem{Tsukamoto+2015oa}
------, ``{Effects of Ohmic and ambipolar diffusion on formation and evolution
  of first cores, protostars, and circumstellar discs},'' \emph{\mnras}, vol.
  452, pp. 278--288, Sep. 2015.

\bibitem{Tsukamoto+2017}
Y.~{Tsukamoto}, S.~{Okuzumi}, K.~{Iwasaki}, M.~N. {Machida}, and S.-i.
  {Inutsuka}, ``{The impact of the Hall effect during cloud core collapse:
  Implications for circumstellar disk evolution},'' \emph{\pasj}, vol.~69,
  p.~95, Dec. 2017.

\bibitem{AllenLiShu2003}
A.~{Allen}, Z.-Y. {Li}, and F.~H. {Shu}, ``{Collapse of Magnetized Singular
  Isothermal Toroids. II. Rotation and Magnetic Braking},'' \emph{\apj}, vol.
  599, pp. 363--379, Dec. 2003.

\bibitem{Larson1969}
R.~B. {Larson}, ``{Numerical calculations of the dynamics of collapsing
  proto-star},'' \emph{\mnras}, vol. 145, p. 271, 1969.

\bibitem{LiMckee1996}
Z.-Y. {Li} and C.~F. {McKee}, ``{Hydromagnetic Accretion Shocks around Low-Mass
  Protostars},'' \emph{\apj}, vol. 464, p. 373, Jun. 1996.

\bibitem{TassisMouschovias2005b}
K.~{Tassis} and T.~C. {Mouschovias}, ``{Magnetically Controlled Spasmodic
  Accretion during Star Formation. II. Results},'' \emph{\apj}, vol. 618, pp.
  783--794, Jan. 2005.

\bibitem{Maclow+1995}
M.-M. {Mac Low}, M.~L. {Norman}, A.~{Konigl}, and M.~{Wardle}, ``{Incorporation
  of ambipolar diffusion into the ZEUS magnetohydrodynamics code},''
  \emph{\apj}, vol. 442, pp. 726--735, Apr. 1995.

\bibitem{ChoiKimWiita2009}
E.~{Choi}, J.~{Kim}, and P.~J. {Wiita}, ``{An Explicit Scheme for Incorporating
  Ambipolar Diffusion in a Magnetohydrodynamics Code},'' \emph{\apjs}, vol.
  181, pp. 413--420, Apr. 2009.

\bibitem{MarchandCommerconChabrier2018}
P.~{Marchand}, B.~{Commer{\c c}on}, and G.~{Chabrier}, ``{Impact of the Hall
  effect in star formation and the issue of angular momentum conservation},''
  \emph{\aap}, vol. 619, p. A37, Oct. 2018.

\bibitem{Price2007}
D.~J. {Price}, ``{splash: An Interactive Visualisation Tool for Smoothed
  Particle Hydrodynamics Simulations},'' \emph{\pasa}, vol.~24, pp. 159--173,
  Oct. 2007.

\end{thebibliography}

\end{document}